\documentstyle[11pt]{l-aa}

\begin{document}

\title{A flexible format for exchanging pulsar data}

\author{ D.R.~Lorimer\inst{1}, A.~Jessner\inst{1}, J.H.~Seiradakis\inst{2}, 
         A.G.~Lyne\inst{3}, N.~D'Amico\inst{4}, A.~Athanasopoulos\inst{2},
         K.M.~Xilouris\inst{5}, M.~Kramer\inst{1}, R.~Wielebinski\inst{1}
       }

\offprints{ D.R.~Lorimer (E-mail: dunc@mpifr-bonn.mpg.de)}

\institute{ Max-Planck-Institut f\"ur Radioastronomie, Auf dem H\"ugel 69,
             D-53121 Bonn, Germany 
\and        University of Thessaloniki, Department of Physics, Section of 
             Astrophysics, Astronomy and Mechanics, GR--54006, Greece
\and        University of Manchester, Nuffield Radio Astronomy Laboratories,
             Jodrell Bank, Macclesfield, Cheshire, SK11 9DL, UK
\and        Osservatorio Astronomico di Bologna, via Zamboni 33, 40126
             Bologna, Italy
\and        National Astronomy and Ionospheric Center, Arecibo Observatory,
             P.O.~Box 995, Arecibo, Puerto Rico 00613
           }

\thesaurus{08.16.6, 04.01.1}

\date{Received  ; accepted }

\maketitle
\markboth{Lorimer et al.: A flexible pulsar data--exchange format}{}

\begin{abstract}

We describe a data format currently in use amongst European
institutions for exchanging and archiving pulsar data. The format is
designed to be as flexible as possible with regard to present and
future compatibility with different operating systems. One application
of the common format is simultaneous multi-frequency observations of
single pulses. A data archive containing over 2500 pulse profiles
stored in this format is now available via the {\it Internet},
together with a small suite of computer programs that can read, write
and display the data.


\end{abstract}

\section{Introduction}

The {\bf E}uropean {\bf P}ulsar {\bf N}etwork (``{\bf EPN}'') is an
association of European astrophysical research institutes that
co--operate in the subject of pulsar research. All institutes have up
until now developed their own individual hardware and software facilities
tailored to their own requirements and will, of course, continue to
do so in future. Contact and co--operation has always existed between
the scientists of the member institutes and outside, but the lack of a
common standard format for pulsar data has hampered previous
collaborative research efforts.

In this paper, we describe a flexible format that we have developed
for exchanging data between EPN pulsar groups. The format has some
generic similarities to the widely used FITS format (Wells et al.~1981)
but has been designed to meet the specific needs of the
EPN.  The format has proved so successful that we now advocate its use
as a useful world--wide utility for pulsar data exchange. To aid
implementation of the format, we have written a suite of freely
available {\it Fortran--77} sub--routines which can be easily
incorporated within existing software to read and write data in this
format. Astrophysical applications of such a format currently being
pursued by EPN groups include the establishment of a data bank of
pulse profiles as well as simultaneous observations of pulsars by
several European observatories.

\section{The EPN format}

The underlying principles of the format result from a number of
requirements. This was essentially a balance between the need for
efficient data storage and providing sufficient information about the
data for potential users. Specifically, the following requirements
had to be met:

\begin{itemize}

\item {\bf Operating system independence:}
To make the data format as portable as possible between present and
future operating systems, we have opted to use only ASCII--data
throughout.  We have arrange these data so that words are aligned
over 80-byte boundaries, this simplifies inspection and printing of
the files.

\item {\bf Completeness:} The data should contain all information for
the identification of the source and the observing circumstances
useful for further analyses of the data by others.

\item {\bf Compactness:} Descriptive information should not dominate 
the format. The measured values that form the bulk of a
block of data are given as scaled four-character hexadecimal numbers,
giving a dynamic range of up to 65536:1.

\item {\bf Versatility:} The format should be suitable for continuously sampled
multi-channel filterbank search data, synchronous integrated and single--pulse
data as well as processed data. In addition, we have designed the format, so
that it can be used for observations of pulsars outside the radio regime {\it i.e.}
variable units for the observing frequency and bandwidth, as well as topocentric
telescope coordinates which are time variable for satellite observatories.
Space is left for more descriptors, future adaptations and expansions.

\item {\bf Simplicity and ease of access:} We describe a data format 
consisting of a 
standardised fixed-length header with a variable length data structure 
attached to it. The header fully describes the structure of the data, which
is not changed within one file but can vary between files. 
In this way, it is possible to calculate the length of a data block within 
each file after reading its header. The file can then be opened for random
access with fixed block length, faster than a sequential read.
\end{itemize} 

Many of the above mentioned requirements were already met by a
format in use at Jodrell Bank to which we made suitable
modifications and extensions to make it more flexible.  

\begin{figure}
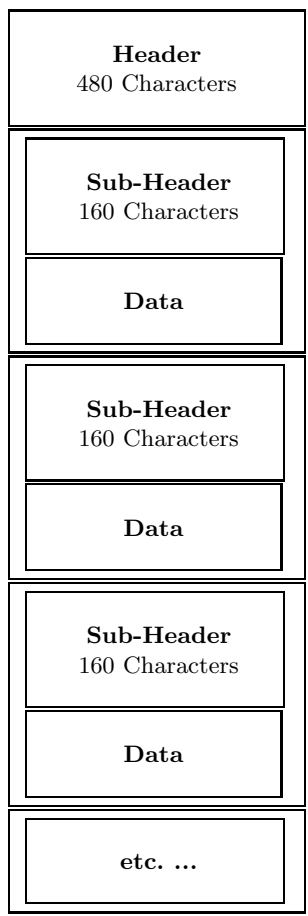

\small
\begin{center}
\begin{minipage}{3.8cm}
\fbox{
 \parbox{3.5cm}{\begin{center} {\bf  Header } \\  480 Characters \end{center}} 
      }
\fbox{
 \begin{minipage}{3.5cm}
  \fbox{ \parbox{3.1cm}{\begin{center} {\bf Sub-Header } \\ 160 Characters \end{center} }} 
  \fbox{ \parbox{2.96cm}{\begin{center}  {\bf Data } \\   \end{center} } }
 \end{minipage}
       }
\fbox{
 \begin{minipage}{3.5cm}
  \fbox{ \parbox{3.1cm}{\begin{center} {\bf Sub-Header } \\ 160 Characters \end{center} }} 
  \fbox{ \parbox{2.96cm}{\begin{center}  {\bf Data } \\   \end{center} } }
 \end{minipage}
       }
\fbox{
 \begin{minipage}{3.5cm}
  \fbox{ \parbox{3.1cm}{\begin{center} {\bf Sub-Header } \\ 160 Characters \end{center} }} 
  \fbox{ \parbox{2.96cm}{\begin{center}  {\bf Data } \\  \end{center} } }
 \end{minipage}
       } 
\fbox{
 \begin{minipage}{3.5cm}
  \fbox{ \parbox{3.1cm}{\begin{center} {\bf etc. ...}  \end{center} }} 
 \end{minipage}
} 
\end{minipage}
\end{center}
\caption{Schematic representation of an EPN data block.}

\end{figure}

Each EPN file consists of one or more EPN blocks.  The basic structure
of an EPN block is shown in Fig. 1.  Each file has a common fixed
length {\it header} followed by a number of individual {\it data
streams} of equal length. The header describes the data, containing
information on the pulsar itself, the observing system used to make
the observation as well as some free-form information about the
processing history of the data. The onus is on the site--specific
conversion process to ensure correct conversion to the standardised
entries and reference to common catalogues (e.g.~the Taylor et al.~1993
\nocite{tml93} catalogue of pulsar
parameters).  The full list of header variables is given in Tables 1
and 2.

The data streams themselves may be outputs of different
polarisation channels, or individual channels (bands) of a filterbank
or a combination thereof. In total, there may be $N_{\rm freq}$ data
streams of i.e. different frequencies for each polarisation.  Each
data stream starts with a small, fixed length sub-header in front of
the actual data values.  The number of data streams and their length may
vary between different EPN files, but is constant within each file.  A
character field and an ordinal number is provided for each stream for
its identification. 
 
\begin{table*}
\begin{center}
\footnotesize
\begin{tabular}{|rcccp{8cm}|}
\hline
Position & Name & Format & Unit & Comment \\
\hline 
\hline
1   &  version &  A8 & &  EPN + version of format (presently EPN05.00)\\
9   &  counter &  I4 & &  No. of records contained in this data block\\
13  &  history & A68  &  &comments and history of the data \\
\hline
81    & jname &  A12 & &  pulsar jname \\
93   & name & A12 &  & common name              \\
105   & $P_{\rm bar}$ &  F16.12 & s & current barycentric period\\
121    & DM      &  F8.3   & pc cm$^{-3}$& dispersion measure\\
129   & RM      &  F10.3  & rad m$^{-2}$ & rotation measure \\
139  & CATREF &  A6   & & pulsar parameter catalogue in use \\
145   & BIBREF    &  A8     & & bibliographical reference key (or observer's name) \\
153   &         & 8X   &  &  blank space free for future expansion \\
\hline
161   & $\alpha_{2000}$  & I2,I2,F6.3 & hhmmss& right ascension of source \\
171   & $\delta_{2000}$ & I3,I2,F6.3 & ddmmss& declination of source\\
182  & telname   & A8  & & name of the observing telescope (site) \\
190  & EPOCH  & F10.3 & day & modified Julian date of observation \\
200  & OPOS   & F8.3  & degrees & position angle of telescope \\
208  & PAFLAG & A1    &  &  A = absolute polarisation position angle, else undefined\\
209  & TIMFLAG & A1   &  &  A = absolute time stamps (UTC), else undefined \\
210 &         & 31X   &  &  blank space free for future expansion \\
\hline
241  & $x_{\rm tel}$& F17.5& m & topocentric X rectangular position of telescope \\
258  & $y_{\rm tel}$& F17.5& m & topocentric Y rectangular position of telescope \\
275  & $z_{\rm tel}$& F17.5& m & topocentric Z rectangular position of telescope \\
292 &         & 29X   &  &  blank space free for future expansion \\
\hline
321  & CDATE & I2,I2,I4 & d m y & creation/modification date of the dataset \\
329  & SCANNO & I4 &  & sequence number of the observation \\
333  & SUBSCAN & I4 &  &sub--sequence number of the observation \\
337  & $N_{\rm pol}$ & I2 &  & number of polarisations observed \\ 
339  & $N_{\rm freq}$ & I4 & & number of frequency bands per polarisation \\
343  & $N_{\rm bin}$ & I4  & & number of phase bins per frequency (1-9999) \\
347  & $t_{\rm bin}$ & F12.6 & $\mu$s & duration (sampling interval) of a phase bin \\
359  & $t_{\rm res}$ & F12.6 & $\mu$s & temporal resolution of the data \\
371  & $N_{\rm int}$ & I6  & & number of integrated pulses per block of data \\
377  & $n_{\rm cal}$ & I4 & $t_{\rm bin}$ & bin number for start of calibration signal\\
381  & $l_{\rm cal}$ & I4 & $t_{\rm bin}$ & length of calibration signal \\
385  & FLUXFLAG      & A1 &    & F = data are flux calibrated in mJy, else undefined \\
386  & & 15X  & & blank space free for future expansion \\
\hline
401 &         & 80X   &  &  blank space free for future expansion \\
\hline
\end{tabular}
\normalsize
\caption{A description of the EPN format variables.}
\end{center}
\end{table*}

\begin{table*}
\begin{center}
\footnotesize
\begin{tabular}{|rcccp{8cm}|}
\hline
Position & Name & Format & Unit & Comment \\
\hline 
\hline
481  & IDfield &  A8 & &  type of data stream (I,Q,U,V etc.) \\
489  & $n_{\rm band}$& I4 & &  ordinal number of current stream  \\
493  & $n_{\rm avg}$ & I4 & & number of streams averaged into the current one \\
497  & $ f_0$ & F12.8 &  & effective centre sky frequency of this stream\\
509  & $ U_f$ & A8 &  & unit of $f_0$ \\
517  & $ \Delta f $&  F12.6 &  & effective band width \\
529  & $ U_{\Delta} $&  A8 &  & unit of $\Delta f$\\
537  & $ t_{\rm start} $ & F17.5 & $\mu$s& time of first phase bin w.r.t. EPOCH \\
554  &               & 7X    &        & blank space free for future expansion\\
\hline
561  & SCALE  & E12.6 &  & scale factor for the data\\
573  & OFFSET & E12.6 &  & offset to be added to the data \\
585  & RMS    & E12.6 &  & rms for this data stream\\
597   & $P_{\rm app}$ &  F16.12 & s & apparent period at time of first phase bin\\
613  &     &  28X    &       & blank space free for future expansion\\
\hline
641 & Data(1)& I4 & & scaled data for first bin \\
$ 4 (N_{\rm bin}-1)+641$& Data($N_{\rm bin}$) & I4 &  & 
data for last bin of stream\\
\hline 
$640 + N_{\rm records}*80$& & & & end of first stream, \\
                          & & & & $N_{\rm records} ={\rm INT}( N_{\rm bin}\cdot 0.05)+\Theta( (4 N_{\rm bin} {\rm mod}\ 80) -1)^* $ \\
\hline
\end{tabular}
\normalsize
\caption{The sub-header variables. $^*\Theta(x)$ is the 
Heaviside--function: $=1$ if $x\ge 0$ and $=0$ elsewhere.}
\end{center}
\end{table*}

\section{Simultaneous observations of single pulses}

Pulsars are, in general, very weak sources, typically requiring the
addition of several thousand individual pulses with a large radio
telescope equipped with sensitive receivers in order to attain a
sufficiently large signal-to-noise ratio.  The brightest pulsars are,
however, strong enough so that individual pulses can be
observed. These pulses are known to exhibit great variety in
morphology and polarimetric properties from one pulse to the next
(see for example Lyne \& Smith 1990). \nocite{ls90}
It is presently unclear whether the same features in the individual
pulses are present at different observing frequencies.  One of the
current research topics being carried out by EPN is a
multi-frequency study of single pulses. The project requires the
pulses observed at different telescopes to be time-aligned and thus
the format described above has an ideal application in this
project. After conversion of the data into this format, the time
alignment of the pulses and subsequent statistical analyses is a
relatively straight-forward procedure.  As an example, a set of pulses
from PSR B0329+54 observed simultaneously at Bologna (410 MHz),
Jodrell Bank (1.404 GHz) and Effelsberg (4.850 GHz) are shown in
Fig. 2. The pulses show a remarkable similarity at these three
frequencies, although counter examples are also observed. Full results
of this study will be published shortly.

\begin{figure*}
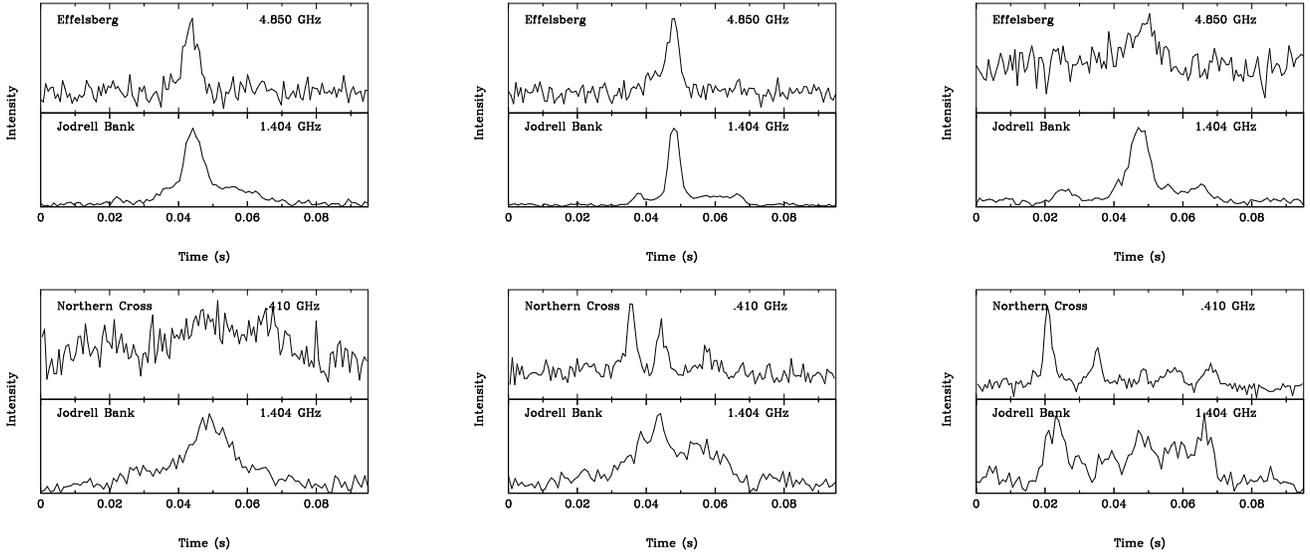

\setlength{\unitlength}{1in}
\begin{picture}(0,2.7)
\put(-.2,6.5){\includegraphics{0329je2.fig}}
\put(-.2,5.0){\includegraphics{0329jb2.fig}}
\end{picture}
\caption[]
{
Time-aligned single pulses for PSR B0329+54 observed simultaneously
at Effelsberg (4.850 GHz) and Jodrell Bank (1.404 GHz) shown in
the upper panel and at Bologna (410 MHz) and Jodrell Bank (1.404 GHz)
in the lower panel. The data were processed using the EPN format.
}
\end{figure*}

\section{The EPN pulse profile archive}

As well as being used for data interchange between EPN members, the
common format forms the basis of a pulse profile archive presently
being maintained at the Max-Planck-Institut f\"ur Radioastronomie in
Bonn. The idea of the archive is to build up a useful collection of
pulse profiles which anybody with access to the {\it Internet} can use.
Presently, around 2500 pulse profiles are stored in this format.
The profiles themselves have usually already been, or about to be, 
published
so that full credit for any subsequent use via the database can
go to the contributing authors. The archive has the following
URL address: {\it http://www.mpifr-bonn.mpg.de/pulsar/data/}.
Authors are encouraged to make their data available to this archive
and should contact Duncan Lorimer (Email: {\it dunc@mpifr-bonn.mpg.de}) 
if they wish to do this.

\acknowledgements 

We thank the referee, Dick Manchester, for useful comments on an earlier
version of the manuscript. The EPN was funded under Brussels Human Capital and
Mobility grant number CHRX-CT94-0622.

\appendix
\section{Format Compatible Software}

To incorporate the capability to read and write data in this format
within existing analysis software, a simple routine exists which can
read and write data in this format. In addition, we have written some
sample programs which can plot the data and display the header
parameters.  The software are written in {\it Fortran---77} and have
been packaged into a single UNIX tar file which is freely available
via the {\it Internet}. To down-load the package, log into the
anonymous ftp area: {\bf ftp.mpifr-bonn.mpg.de}, with the username
{\bf anonymous} using your complete E--mail address as the
password. Once logged in, issue the following commands:

\begin{verbatim}
cd pub/pulsar
binary
get epnsoft.tar.gz
\end{verbatim}

\noindent
Alternatively, the file can be down-loaded from the EPN {\it Internet}
home-page: {\it http://www.mpifr-bonn.mpg.de/pulsar/epn/}.

\bigskip
\noindent
To uncompress and extract the contents of the tar file on a UNIX
operating system, issue the commands:

\begin{verbatim}
gunzip epnsoft.tar.gz
tar xvf epnsoft.tar
\end{verbatim}

\noindent
The present package contains some sample data and two example programs ---
``plotepn'' and ``viewepn'' which plot and view EPN files respectively.
The ASCII file {\bf 00README} in this packages gives further details of the 
software and how to use it.

\begin{thebibliography}{}

\bibitem[Lyne \& Smith<1990>]{ls90}
Lyne~A.~G., Smith~F.~G., 1990, Pulsar Astronomy.
\newblock Cambridge University Press

\bibitem[Taylor, Manchester \& Lyne<1993>]{tml93}
Taylor~J.~H., Manchester~R.~N., Lyne~A.~G., 1993, ApJS, 88, 529

\bibitem[Wells, Greisen \& Harten<1981>]{wgh81}
Wells~D.~C., Greisen~E.~W., Harten~R.~H., 1981, A\&A Supp. Ser, 44, 363

\end{thebibliography}
\end{document}